\documentclass[aps,prb,reprint,letterpaper,showpacs,twocolumn,nofootinbib,superscriptaddress,floatfix]{revtex4-1}
\usepackage[UKenglish]{babel}
\usepackage{graphicx}
\usepackage[pdftex                 
           ,pagebackref=false   
           ,colorlinks=true     
           ]{hyperref}
\hypersetup{linkcolor= blue,  
            citecolor= red,    
            urlcolor=black}   

\usepackage{amsmath}
\usepackage{mathtools}
\usepackage{eufrak}
\usepackage{times}
\usepackage{bm}
\usepackage{soul,color}
\usepackage{bbold}

\bibpunct{[}{]}{,}{n}{}{}

\newcommand*{\fleur}{\texttt{FLEUR}}
\newcommand*{\wprog}{\textsc{wannier}{\footnotesize{90}}}
\newcommand*{\vn}[1]{\bm{{#1}}}
\newcommand*{\ut}[1]{\mathrm{#1}}

\begin{document}

\date{5th November 2019}
\begin{abstract}
Integrating topologically stabilized magnetic textures such as skyrmions as nanoscale information carriers into future technologies requires the reliable control by electric currents. Here, we uncover that the relevant skyrmion Hall effect, which describes the deflection of moving skyrmions from the current flow direction, acquires important corrections owing to anisotropic spin-orbit torques that alter the dynamics of topological spin structures. Thereby, we propose a viable means for manipulating the current-induced motion of skyrmions and antiskyrmions. Based on these insights, we demonstrate by first-principles calculations and symmetry arguments that the motion of spin textures can be tailored by materials design in magnetic multilayers of Ir/Co/Pt and Au/Co/Pt. Our work advances the understanding of the current-induced dynamics of these magnetic textures, which underlies a plethora of memory and logic applications.

\end{abstract}

 \title{Engineering the dynamics of topological spin textures by anisotropic spin-orbit torques}
 \author{J.-P. Hanke}
 \email{j.hanke@fz-juelich.de}
 \affiliation{Peter Gr\"unberg Institut and Institute for Advanced Simulation,\\Forschungszentrum J\"ulich and JARA, 52425 J\"ulich, Germany}
 \affiliation {Institute of Physics, Johannes Gutenberg University Mainz, 55099 Mainz, Germany} 
 \author{F. Freimuth}
 \affiliation{Peter Gr\"unberg Institut and Institute for Advanced Simulation,\\Forschungszentrum J\"ulich and JARA, 52425 J\"ulich, Germany} 
 \author{B. Dup{\'e}}
 \affiliation{Fonds de la Recherche Scientifique -- FNRS}
 \affiliation{Nanomat/Q-mat/CESAM, Universit\'e de Li\`ege, B-4000 Sart Tilman, Belgium} 
 \affiliation {Institute of Physics, Johannes Gutenberg University Mainz, 55099 Mainz, Germany} 
 \author{J. Sinova}
 \author{M. Kl\"aui}
 \affiliation {Institute of Physics, Johannes Gutenberg University Mainz, 55099 Mainz, Germany}
 \author{Y. Mokrousov}
 \email{y.mokrousov@fz-juelich.de}
 \affiliation{Peter Gr\"unberg Institut and Institute for Advanced Simulation,\\Forschungszentrum J\"ulich and JARA, 52425 J\"ulich, Germany}
 \affiliation {Institute of Physics, Johannes Gutenberg University Mainz, 55099 Mainz, Germany} 
 \maketitle

\section{Introduction}
Magnetic skyrmions are topologically stabilized spin textures that hold bright promises as robust processing units in innovative information technologies. In particular, the prospects of using these non-collinear magnetic structures in efficient racetrack memories~\cite{Fert2013} or for brain-inspired computing~\cite{Bourianoff2018,Zazvorka2019} have sparked tremendous interest in the research field.  Ever since, generating these localized spin structures, moving them at high speeds, and detecting them by electrical means constitute the key objectives for realizing competitive skyrmion-based devices. Only recently, it has been experimentally demonstrated that individual skyrmions can form at room temperature~\cite{Woo2016,MoreauLuchaire2016,Everschor2018} due to an enhanced thermal stability in carefully designed magnetic superlattices, for instance, of Pt/CoFeB/MgO or Ir/Co/Pt. In these systems combining spatial inversion asymmetry with strong spin-orbit coupling, the stabilization of topological spin textures originates from the Dzyaloshinskii-Moriya interaction (DMI)~\cite{Dzyaloshinsky1958,Moriya1960}. Its interfacial nature renders this chiral exchange interaction tunable in the magnetic multilayers via appropriate materials engineering~\cite{Hrabec2014,Yang2015a,Hanke2018}, which facilitates the versatile control over static texture properties including skyrmion radius and chirality~\cite{Camosi2017,Hoffmann2017}.

The challenge of displacing these chiral magnetic structures by purely electrical means is usually addressed by exploiting the phenomenon of spin-orbit torques (SOTs)~\cite{Manchon2019}. These torques contain two qualitatively distinct contributions, namely, a field-like term $\vn T^\ut{FL}$ and an antidamping term $\vn T^\ut{AD}$, both of which root in relativistic spin-orbit effects in systems that lack spatial inversion symmetry. Specifically, charge currents can trigger an interfacial spin polarization in these systems, exerting SOTs on the local magnetization~\cite{Chernyshov2009,Miron2010,Miron2011,Garello2013,Freimuth2014a}, which facilitates switching of ferro- and antiferromagnets~\cite{Miron2011a,Liu2012,Wadley2016} as well as moving of domain walls and skyrmions~\cite{Tomasello2014,Jiang2015,Woo2016,Litzius2017}. However, owing to their non-trivial topology in two-dimensional real space, skyrmions move under an angle $\theta_\ut{sk}$ with respect to the line of the applied current~\cite{Jiang2015,Litzius2017}. This so-called skyrmion Hall effect obstructs the immediate use of skyrmionic spin structures in racetrack devices, e.g., for logic functionalities, as they suffer intrinsically from deflections. Shaping the complex trajectory of topological spin textures fundamentally relies on our microscopic understanding of the interplay between spin topology, damping, and current-induced SOTs~\cite{Ritzmann2018}. While the Thiele equation of motion~\cite{Thiele1973,Clarke2008} accounts for this interplay, the corresponding treatment led to the common perception that a whole class of SOTs -- the field-like torques -- is irrelevant for describing the dynamical properties of rigid skyrmions as well as the skyrmion Hall effect. 

In this work, we demonstrate that this widely accepted picture needs to be extended owing to the non-trivial form of SOTs in magnetic multilayers. While the current-induced dynamics of skyrmions was previously explained based on deformations of the spin texture~\cite{Litzius2017}, here, we promote a distinct mechanism to activate field-like SOTs. We show that the anisotropy of these torques plays an important role for predicting and interpreting the dynamics of topological spin structures. Specifically, we uncover that the coupling of magnetic textures to field-like torques which are higher order in the local magnetization can manifest in large corrections to the skyrmion Hall effect. Using first-principles calculations and symmetry arguments, we quantify the relevance of these modifications for the dynamical properties of skyrmions and antiskyrmions in layered magnetic films of Ir/Co/Pt and Au/Co/Pt, as described by the Thiele equation. Our findings outline a new perspective for controlling the skyrmion Hall effect and the motion of skyrmions or antiskyrmions in multilayer systems by engineering the anisotropy of the SOTs.

The article is structured as follows. In Sec.~\ref{sec:methods}, we review the Thiele equation of current-induced skyrmion motion, elucidate the symmetry-allowed form of SOTs based on effective magnetic fields, and provide the details of our first-principles calculations. The resulting anisotropies of the torques in the magnetic trilayers are presented in Sec.~\ref{sec:results}, where we discuss also their imprint on the forces and skyrmion Hall angle of topological spin textures. Section~\ref{sec:conclusions} concludes this work.

\section{Methods}\label{sec:methods}
\subsection{Thiele equation of skyrmion motion}
Treating skyrmions as rigid objects in a ferromagnetic background, the Thiele equation~\cite{Thiele1973,Clarke2008} describes the dynamical properties of unit-vector magnetization fields $\vn m(\vn r)$ with non-trivial topology in two-dimensional real space:
\begin{equation}
 ( Q\mathcal A - \mathcal D) \vn v = \vn F \, ,
\label{eq:thiele}
\end{equation}
where $\vn v=(v_x,v_y)$ is the velocity of the skyrmions. The gyromagnetic term $Q\mathcal A \vn v$ in Eq.~\eqref{eq:thiele} is mediated by the antisymmetric tensor $\mathcal A_{ij}=\epsilon_{ij}$ with the Levi-Civita symbol $\epsilon_{ij}$ and the integer topological charge
\begin{equation}
    Q=(1/4\pi)\int_{\mathbb{R}^2} \vn m \cdot (\partial_x \vn m \times \partial_y \vn m) \, \mathrm d\vn r \, .
\end{equation}
Additionally, the magnetization dynamics are affected by the dissipative coupling $-\mathcal D\vn v$, which stems from
\begin{equation}
\mathcal D_{ij}=(\alpha_\ut{G}/4\pi)\int_{\mathbb{R}^2} (\partial_i \vn m \cdot \partial_j \vn m) \, \mathrm d\vn r \, ,
\label{eq:dissipation}
\end{equation}
where $\alpha_\ut{G}$ is the Gilbert damping, and $\partial_i \vn m=\partial \vn m / \partial r_i$ denotes the spatial gradient of the spin texture along the $i$th Cartesian direction. While gyromagnetic and dissipative terms root manifestly in the local variations of $\vn m(\vn r)$ in real space, the generalized force $\vn F=(F_x,F_y)$ on the right-hand side of Eq.~\eqref{eq:thiele} reflects the interaction of current-induced torques $\vn T(\vn m)$ with the magnetic texture:
\begin{equation}
F_i = \frac{1}{4\pi} \int_{\mathbb{R}^2} \vn T(\vn m) \cdot (\partial_i {\vn m} \times {\vn m}) \, \mathrm d\vn r \, .
\label{eq:force}
\end{equation}
To arrive at this expression, locally, the spin polarization perpendicular to $\vn m$ is expressed as $\vn s_\perp = \vn m \times \vn T(\vn m)$~\cite{Ado2017}.

Using polar coordinates $\vn r=(\rho,\phi)$, we focus in this work on N\'eel skyrmions with a radially symmetric profile as given by $\vn m(\rho,\phi)=[\cos\phi\sin\Theta(\rho),\sin\phi\sin\Theta(\rho),\cos\Theta(\rho)]$ that is characterized by the position-dependent angle~\cite{Romming2015}
\begin{equation}
    \Theta(\rho) = \arcsin\left(\tanh\frac{c-\rho}{w/2}\right) - \arcsin\left(\tanh\frac{c+\rho}{w/2}\right) \, .
    \label{eq:profile}
\end{equation}
Here, the parameter $c$ represents the size of the skyrmion core, and $w$ denotes the width of the domain wall, which we will use below as the intrinsic length scale for the real-space quantities $\rho$ and $c$. A schematic of the radial profile is shown in Fig.~\ref{fig:forces}(a).

\begin{figure*}
    \centering
    \scalebox{0.18}{\includegraphics{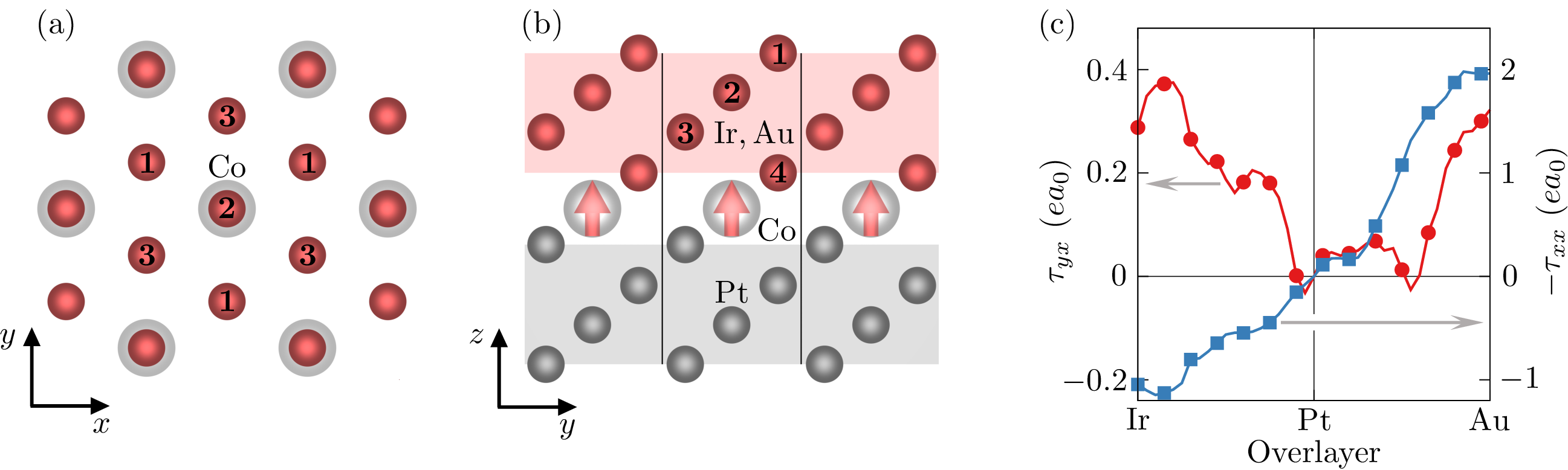}}
    \caption{(a,b)~Top and side views of the considered magnetic films that consist of Pt(111) underlayers (dark gray), a magnetic Co monlayer (light gray), and heavy-metal overlayers (dark red) such as Ir or Au. Bold integers label the different overlayers to highlight the $C_{3v}$ symmetry around the $z$ axis. Red arrows in (b) indicate the direction of the magnetic moment of Co. (c)~Variation of the spin-orbit torkances $\tau_{yx}$ (red circles) and $-\tau_{xx}$ (blue squares) when alloying the overlayers from Ir over Pt to Au in the perpendicularly magnetized system. The effect of alloying is treated within the virtual crystal approximation.}
    \label{fig:structure_and_vca}
\end{figure*}

\subsection{Spin-orbit torques and effective fields}
According to Eq.~\eqref{eq:force}, current-induced torques acting on the magnetization constitute the key information in understanding and predicting the dynamical properties of spin textures based on the Thiele equation. Restricting our analysis here to the case of spin-orbit torques (SOTs), we distinguish two qualitatively distinct contributions to these phenomena, namely, the field-like torque $\vn T^\ut{FL}(\vn m) = \vn m \times \vn H^\ut{FL}_\ut{eff}(\vn m)$ and the antidamping torque $\vn T^\ut{AD}(\vn m) = \vn m \times \vn H^\ut{AD}_\ut{eff}(\vn m)$, which can be related to effective magnetic fields that are routinely used to quantify SOTs in experiment. While the field-like component of the torque changes its sign under magnetization reversal, the antidamping one is invariant, which is also reflected in the microscopic form of the fields $\vn H^\ut{FL}_\ut{eff}$ and $\vn H^\ut{AD}_\ut{eff}$.

All of the symmetry-allowed terms in the magnetization dependence of the field-like and antidamping torques, entering Eq.~\eqref{eq:force}, can be obtained systematically by applying the so-called Neumann's principle~\cite{Voigt1928}. When interpreting the SOTs as linear response to the electric field $\vn E$, the corresponding effective magnetic fields assume the forms
\begin{align}
 H_{\ut{eff},i}^\ut{FL} &= \alpha_{ij} E_j + \alpha_{ijkl} E_j m_k m_l + \ldots \, , \label{eq:neumann_fl} \\
 H_{\ut{eff},i}^\ut{AD} &= \beta_{ijk} E_j m_k  + \beta_{ijkln} E_j m_k m_l m_n + \ldots \, .
 \label{eq:neumann_ad}
\end{align}
Here, summation over repeated indices is implied, and the axial tensors $\alpha$ as well as the polar tensors $\beta$ follow the symmetries of the underlying crystal lattice according to Neumann's principle. While the effective fields for the field-like torques originate from axial tensors of even rank, their antidamping analogues root in the polar tensors of odd rank. Thus, if the electric field is oriented along the $x$ direction within the plane of the two-dimensional film, the effective fields are $\vn H_\ut{eff}^\ut{FL}(\vn m)\propto\hat{\vn e}_y$ and $\vn H_\ut{eff}^\ut{AD}(\vn m)\propto(\vn m \times \hat{\vn e}_y)$ to lowest order. We demonstrate below that non-trivial higher-order terms in the expansions, which go beyond these conventionally used lowest-order expressions, are important as they modify the dynamical properties of topological spin textures.

Depending on the symmetry of a given system, each tensor in Eqs.~\eqref{eq:neumann_fl} and~\eqref{eq:neumann_ad} may contain several free parameters that are specific to the electronic structure of the material. In order to determine these parameters from first principles, we evaluate first the torkance tensor $\tau(\vn m)$, which relates the SOTs to the applied electric field via $T_i(\vn m)=\sum_j \tau_{ij}(\vn m) E_j$~\cite{Freimuth2014a}. The shape of the torkance tensor itself depends on the crystal symmetry and on the magnetization direction. For example, the field-like torkance is proportional to the unit matrix for films with $C_{3v}$ symmetry and perpendicular magnetization, whereas the antidamping torkance amounts to an antisymmetric tensor in this case. Based on linear response theory, the (even) antidamping torkance is given by~\cite{Freimuth2014a}
\begin{equation}
\begin{split}
    \tau_{ij}^\ut{AD} =& \frac{e\hbar}{2\pi N}\sum_{\vn k n}\sum_{m\neq n} \mathrm{Im}[ \langle \psi_{\vn kn} | \mathcal T_i | \psi_{\vn km} \rangle \langle \psi_{\vn km} | v_j | \psi_{\vn kn}\rangle ] \\
    &\times\bigg\{ \frac{\Gamma(\mathcal E_{\vn km} - \mathcal E_{\vn kn}}{[ (\mathcal E_\ut{F} - \mathcal E_{\vn kn})^2 + \Gamma^2 ] [ (\mathcal E_\ut{F} - \mathcal E_{\vn km})^2 + \Gamma^2 ]} \\
    &+ \frac{2\Gamma}{[\mathcal E_{\vn kn} - \mathcal E_{\vn km} ] [(\mathcal E_\ut{F}-\mathcal E_{\vn km})^2 + \Gamma^2 ] } \\
    &+ \frac{2}{(\mathcal E_{\vn kn}-\mathcal E_{\vn km})^2} \mathrm{Im} \ln \frac{\mathcal E_{\vn km} - \mathcal E_\ut{F} - \mathrm i \Gamma}{\mathcal E_{\vn kn} - \mathcal E_\ut{F} - \mathrm i \Gamma} \bigg\} \, ,
 \label{eq:torkance_even} 
 \end{split}
\end{equation}
and the (odd) field-like torkance amounts to~\cite{Freimuth2014a}
\begin{equation}
    \tau_{ij}^\ut{FL} =  \frac{e\hbar}{\pi N} \sum_{\vn knm} \frac{\Gamma^2 \mathrm{Re}[ \langle \psi_{\vn kn}| \mathcal T_i | \psi_{\vn km}\rangle \langle \psi_{\vn km} | v_j | \psi_{\vn kn} \rangle ]}{[ (\mathcal E_\ut{F} - \mathcal E_{\vn kn})^2 + \Gamma^2 ] [ (\mathcal E_\ut{F} - \mathcal E_{\vn km})^2 + \Gamma^2 ] }\, , \label{eq:torkance_odd}
\end{equation}
where $N$ is the number of $\vn k$-points, $|\psi_{\vn kn}\rangle$ is the Bloch state with the band energy $\mathcal E_{\vn kn}$, the Fermi energy is $\mathcal E_\ut{F}$, and $\Gamma$ is a constant band broadening that models the effect of disorder. In addition, $\hbar \vn v = \nabla_{\vn k} H(\vn k)$ and $\vn{\mathcal T} = \vn m \times \nabla_{\vn m} H(\vn k)$ denote velocity and torque operators, respectively, which relate to momentum and magnetization derivatives of the lattice-periodic Hamiltonian. Then, by fitting the analytical forms of the torques due to Eqs.~\eqref{eq:neumann_fl} and~\eqref{eq:neumann_ad} to first-principles data for the corresponding torkances, we extract the material-specific parameters that enter the Neumann's expansion of the effective magnetic fields. Thereby, we can gain microscopic insights into the magnetization dependence of the SOTs, which allows us to assess their roles for the dynamics of rigid spin structures as governed by the Thiele equation~\eqref{eq:thiele}.

\subsection{Details of first-principles calculations}
Using the full-potential linearized augmented-plane-wave (FLAPW) method as implemented in the \fleur{} code~\cite{fleur}, we perform density functional theory calculations of the electronic structure of thin ferromagnetic films. Specifically, we consider hexagonal magnetic trilayers which consist of a Co monolayer between four Pt(111) layers and four alloyed heavy-metal layers that contain Ir or Au, see Fig.~\ref{fig:structure_and_vca}(a,b). The effect of alloying is accounted for by changing the nuclear charges of the overlayers under the constraint of charge neutrality within the virtual crystal approximation (VCA)~\cite{Bellaiche2000}. The structural parameters of Ref.~\cite{Hanke2018} are adopted, and exchange-correlation effects are treated within the generalized gradient approximation~\cite{rpbe}. We choose a muffin-tin radius of $2.23\,a_0$ for Co and $2.29\,a_0$ for all other atom types, where $a_0$ refers to Bohr's radius. The plane-wave cutoff is set to $4.0\,a_0^{-1}$ and spin-orbit coupling is included self-consistently. In the studied ferromagnetic systems with $C_{3v}$ crystal symmetry, the magnetization direction $\vn m=(\cos\varphi\sin\theta,\sin\varphi\sin\theta,\cos\theta)$ is represented in spherical coordinates by the polar angle $\theta$ and the azimuthal angle $\varphi$.

After converging charge and spin densities of the trilayers, we exploit the obtained wave-function information on an equidistant $\vn k$-mesh of $8$$\times$$8$ points for $8$ different values of $\theta$ or $\varphi$ in $[0,2\pi)$ in order to construct higher-dimensional Wannier functions (HDWFs)~\cite{Hanke2015}. Based on a customized version of the \wprog{} code~\cite{Mostofi2014}, we generate a single set of $162$ HDWFs out of $228$ Bloch bands, with the frozen window extending up to $2$\,eV above the Fermi energy. The representation of the Hamiltonian in the basis of HDWFs allows us to efficiently access the full dependence of the SOTs on $\theta$ and $\varphi$ via a generalized Wannier interpolation~\cite{Hanke2015,Hanke2018}. To obtain accurate values for the field-like and antidamping torkances for arbitrary magnetization directions, we sample the momentum Brillouin zone with a dense mesh of $512$$\times$$512$ points to carry out the $\vn k$-summation in Eqs.~\eqref{eq:torkance_even} and~\eqref{eq:torkance_odd}. Furthermore, we employ a broadening of $\Gamma=25\,$meV to model the effect of disorder on the energy bands.

\begin{figure*}
    \centering
    \scalebox{1.0}{\includegraphics{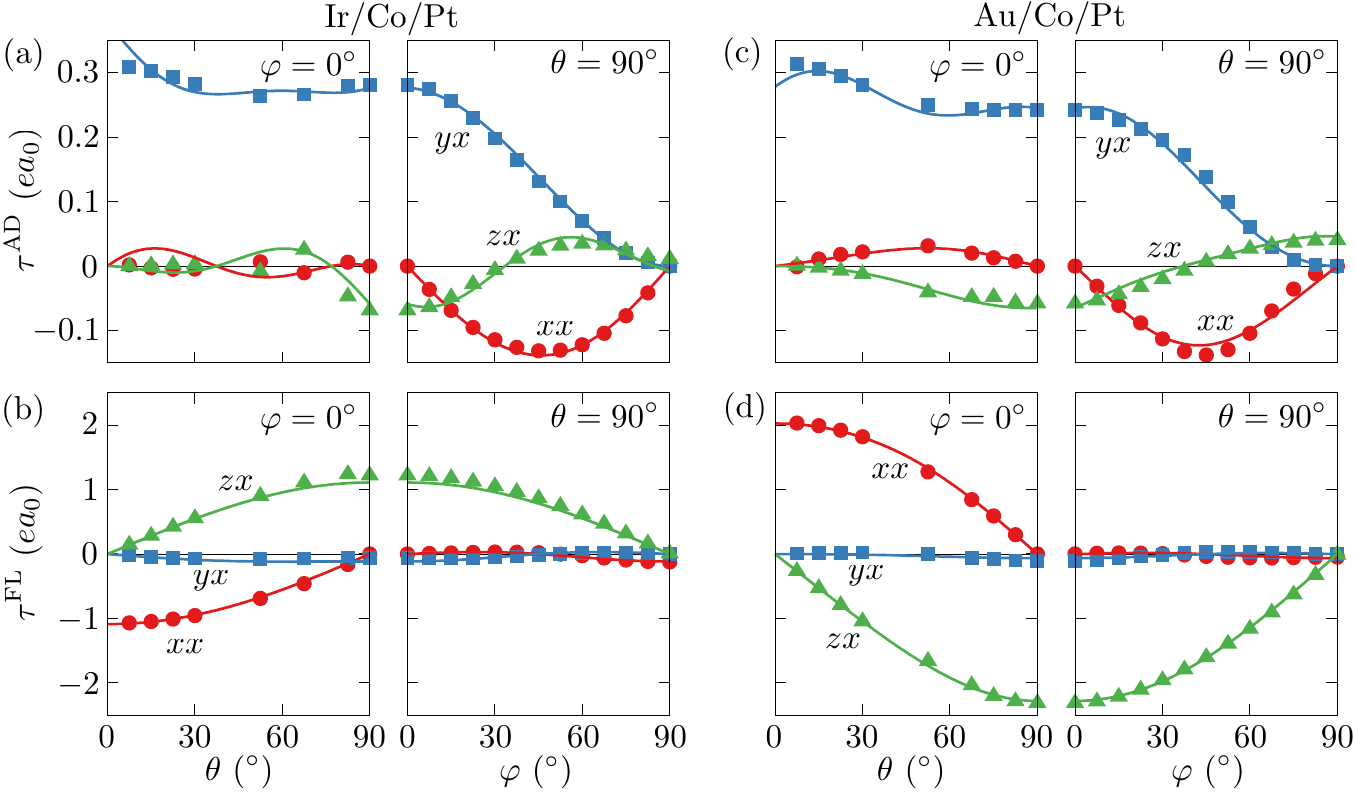}}
    \caption{(a,b)~First-principles dependence of (a)~antidamping and (b)~field-like torkance tensors on the magnetization direction in the ferromagnetic trilayer Ir/Co/Pt. The angles $\theta$ and $\varphi$ are used to characterize the magnetization direction $\vn m=(\cos\varphi\sin\theta,\sin\varphi\sin\theta,\cos\theta)$ in spherical coordinates. Filled circles, squares, and triangles denote data for the $xx$, $yx$, and $zx$ elements of the tensors, respectively, and solid lines represent corresponding fits according to the Neumann expansion. (c,d)~Same as in panels~(a) and~(b) but for Au/Co/Pt.}
    \label{fig:torques}
\end{figure*}

\section{Results}\label{sec:results}
\subsection{Effect of heavy-metal overlayers}
First, we discuss how the field-like and antidamping SOTs are modified as we consider different alloys for the heavy-metal overlayers of the perpendicularly magnetized system with $\theta=\varphi=0^\circ$. Figure~\ref{fig:structure_and_vca}(c) reveals that the antidamping torkance $\tau_{yx}$ is rather susceptible to the corresponding changes of the electronic structure, which directly correlates with the prominent sensitivity of the Dzyaloshinskii-Moriya interaction with respect to alloying~\cite{Hanke2018}. However, the computed values of $\tau_{yx}$ are nearly identical in the stoichiometric Ir/Co/Pt and Au/Co/Pt films, which lack spatial inversion symmetry. In sharp contrast, the larger field-like response $\tau_{xx}$ follows an approximately linear trend with alloy composition, see Fig.~\ref{fig:structure_and_vca}(c), resulting in torkances of similar magnitude but opposite sign in the two stoichiometric cases. We focus in the following discussion on the trilayers Ir/Co/Pt and Au/Co/Pt.

\subsection{Anisotropy of spin-orbit torques}
We begin with elucidating the magnetization dependence of the spin-orbit torkance tensors in the ferromagnetic trilayers for the example of an applied electric field along the $x$ direction. Using the above usual lowest-order expressions for the effective fields $\vn H_\ut{eff}^\ut{FL}$ and $\vn H_\ut{eff}^\ut{AD}$, we expect that the $xx$ element of the odd field-like torkance follows a curve like $\tau_{xx}^\ut{FL}\propto\cos\theta$ if $\varphi=0^\circ$, whereas the even antidamping torkance $\tau_{yx}^\ut{AD}$ should amount to a constant value in this case.

Our first-principles results shown in Fig.~\ref{fig:torques} clearly demonstrate that, in fact, this simple picture does not hold in Ir/Co/Pt and Au/Co/Pt, indicating that higher-order torques which are allowed by symmetry play an important role. In particular, the even torkance is strongly anisotropic with respect to the magnetization direction, Fig.~\ref{fig:torques}(a,c), and its functional form is qualitatively different in the two layered systems owing to distinct spin-orbit hybridizations of the states near the Fermi energy. Analogously, the angular dependence of the odd torkance $\tau_{xx}^\ut{FL}$ in Fig.~\ref{fig:torques}(b,d) departs from the ideal cosine-like shape, irrespective of the considered heavy-metal overlayer which determines the overall sign of the field-like torque. However, while the odd torkance is nearly one order of magnitude larger than its even counterpart, the deviation from the anticipated lowest-order behavior is less prominent than for the antidamping torkance. 

By fitting the corresponding Neumann expansions for the effective fields for the case of $C_{3v}$ crystal symmetry to our complete data, we reveal that higher-order terms (i.e., polar tensors of rank five) are indeed necessary to describe the computed angular dependence of the antidamping SOTs. Specifically, we find that the effective field $\vn H_\ut{eff}^\ut{AD} \propto (0,0,m_x^2m_z)$ constitutes the first important torque correction with a similar value of the corresponding free parameter in the systems Ir/Co/Pt and Au/Co/Pt. For example, this qualitatively distinct higher-order correction amounts to as much as $40\%$ of the leading order in Ir/Co/Pt, which underlines its relevance for understanding the magnetization dependence of antidamping SOTs. In the case of the field-like torques, the same fitting procedure predicts that the effective field $\vn H_\ut{eff}^\ut{FL} \propto \hat{\vn e}_y$ describes the SOT data to first approximation rather well. However, a more detailed analysis uncovers that higher-order contributions beyond this leading-order term are present in both systems, as encoded in axial tensors of rank four in Eq.~\eqref{eq:neumann_fl} that give rise to SOT fields like $\vn H_\ut{eff}^\ut{FL} \propto (2m_xm_y,-3m_x^2-m_y^2,0)$ and $\vn H_\ut{eff}^\ut{FL}\propto (0,-m_z^2,0)$. The anisotropic field-like torques which result from these fields will play a key role for the dynamical properties of topological spin textures.

\begin{figure}
    \centering
    \scalebox{0.18}{\includegraphics{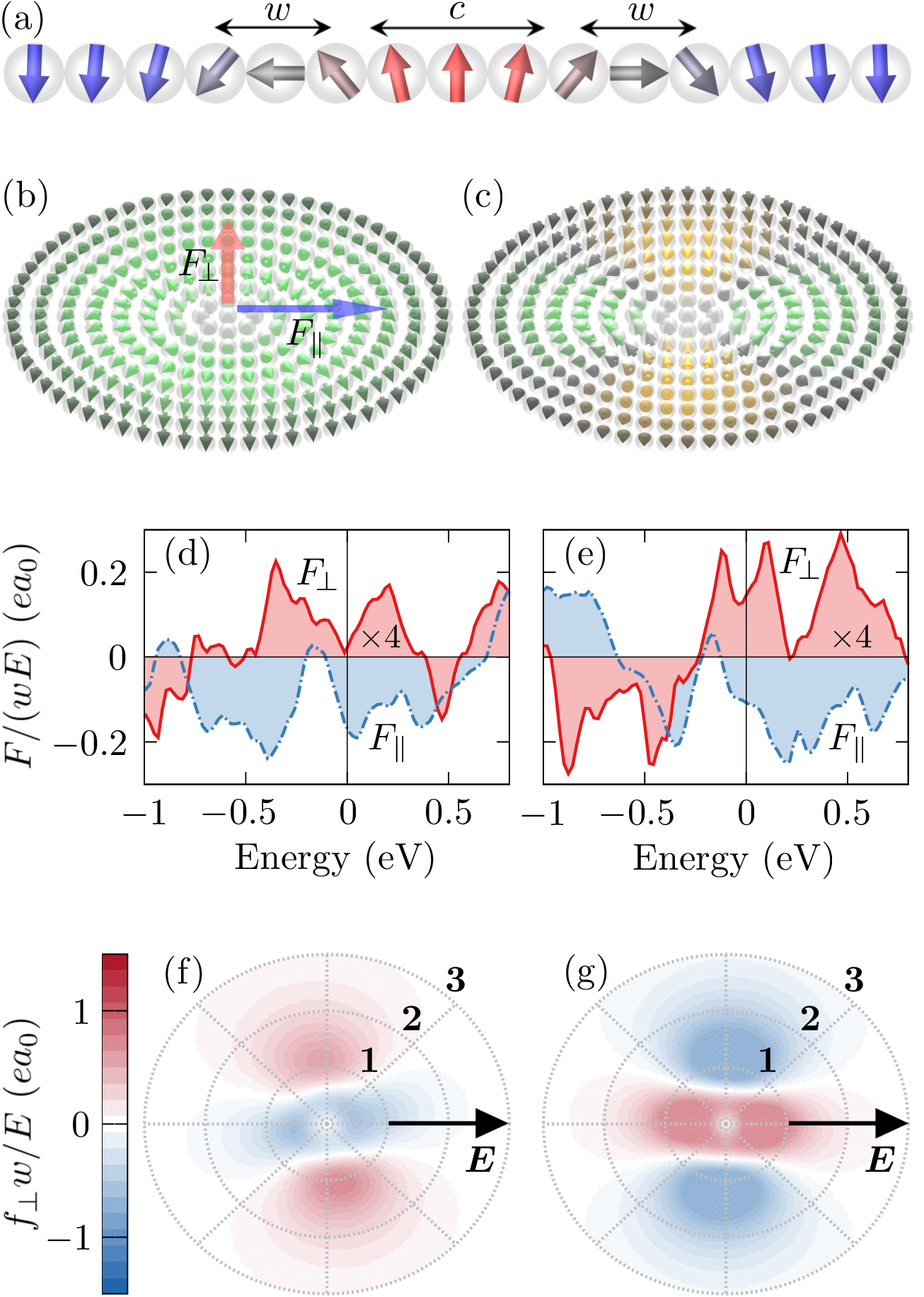}}
    \caption{(a)~Radially symmetric N\'eel-type skyrmion profile with core size $c$ and width $w$ of the domain wall. Colors encode the out-of-plane component of the local magnetization. (b,c)~Skyrmions and antiskyrmions are subject to current-induced forces that act parallel ($F_\parallel)$ or perpendicular ($F_\perp$) to the current flow direction. Green and yellow colors correspond to clockwise and counter-clockwise rotational senses along the radial direction, respectively, and the brightness denotes the out-of-plane spin component. (d,e)~Forces $F_\parallel$ (dashed blue line) and $F_\perp$ (solid red line), per electric field $E$, on the radially symmetric N\'eel skyrmion in the magnetic trilayers (d)~Ir/Co/Pt and (e)~Au/Co/Pt. The position of the Fermi energy is varied, the value of the force $F_\perp$ perpendicular to the electric-field direction is scaled by a factor of four, and $c=w/2$. (f,g)~Real-space distribution of the force density $f_\perp$ in (f)~Ir/Co/Pt and (g)~Au/Co/Pt for the self-consistent value of the Fermi energy. Bold integers as well as circles of increasing radius indicate the radial distance $\rho$ in units of $w$, the skyrmion core is located in the center, and the arrow points along the electric field $\vn E$.}
    \label{fig:forces}
\end{figure}

\subsection{Generalized forces on spin textures}
Next, we turn to the Thiele equation of motion and evaluate the generalized forces therein which originate from the current-induced torques according to Eq.~\eqref{eq:force}. Leaving aside the effect of spin-transfer torques for future work, here, we aim to uncover the fundamental signatures of SOTs for the dynamical properties of topological spin structures. For this purpose, we assume that the electronic structure of non-trivial spin textures follows locally the one of the ferromagnetic state with a given direction $\vn m$ such that we can exploit our first-principles SOT data to calculate the torque-driven forces. In particular, we focus on the coupling of SOTs to radially symmetric N\'eel-type skyrmions following the magnetization profile given by Eq.~\eqref{eq:profile} with $c=w/2$, see also Fig.~\ref{fig:forces}(a).

Figure~\ref{fig:forces} summarizes our first-principles results for the generalized in-plane forces which act parallel ($F_\parallel$) or perpendicular ($F_\perp$) to the applied electric field. The calculated dependence of these forces on the position of the Fermi energy in Fig.~\ref{fig:forces}(d,e) reveals that they are susceptible to changes of the local electronic structure. Thus, doping of the heavy-metal overlayer provides a suitable means for tailoring the magnitude and the sign of these dynamical forces. In addition, we remark that the characteristic angular behavior of the antidamping torques manifests exclusively in a parallel force component for symmetric spin textures, whereas the force $F_\perp$ is completely determined by the field-like SOTs.

Although the magnitude of the field-like torques is generally larger (see Fig.~\ref{fig:torques}), the resulting forces are a factor of four smaller than those due to the antidamping SOTs. This can be understood by analyzing in detail the form of Eq.~\eqref{eq:force} when taking into account the expansions in Eqs.~\eqref{eq:neumann_fl} and~\eqref{eq:neumann_ad}. While the effective field $\vn H_\ut{eff}^\ut{AD}$ generates a finite force already at its lowest order, it is straightforward to show that the field-like SOT mediated by $\vn H_\ut{eff}^\ut{FL}\propto \hat{\vn e}_y$ results in no force on the rigid symmetric spin texture. As a consequence, the less prominent higher-order terms of $\vn H_\ut{eff}^\ut{FL}$ constitute the only sources for the non-zero transverse force $F_\perp$ in Fig.~\ref{fig:forces}(d,e) due to the field-like torques. Therefore, by controlling the anisotropy of the corresponding effective SOT fields, we could engineer the magnitude of the force that acts perpendicular to the current line. Alternatively, we note that the inoperative leading order of the field-like torques would become active again if the spin texture is deformed as is often observed in experiment (see, for example, Ref.~\cite{Litzius2017}).

Based on our first-principles results shown in Fig.~\ref{fig:forces}(d,e), we identified the polar and axial effective fields that yield the most dominant contributions to the torque-driven forces. In both systems Ir/Co/Pt and Au/Co/Pt, the largest terms due to the field-like SOTs originate from the effective fields $\vn H_\ut{eff}^\ut{FL}$ that behave like $(2 m_x m_y,-3 m_x^2-m_y^2,0)$, $(0,-m_z^2,0)$, and $(m_x m_y,m_y^2,0)$, given in descending order of relevance for the forces. In the case of the antidamping SOTs, similarly, we find that the effective fields $\vn H_\ut{eff}^\ut{AD}$ which are proportional to $(m_z,0,0)$, $(0,0,m_x)$, and $(0,0,m_x^2 m_z)$ constitute the most important parts of the forces in the considered films.

We present in Fig.~\ref{fig:forces}(f,g) the real-space distributions of the density $f_\perp$ as the integrand in Eq.~\eqref{eq:force} that sums up to the transverse generalized force $F_\perp$. The panels uncover that this force density, which reflects the non-trivial angular dependence of the anisotropic field-like SOTs, is distributed asymmetrically with a generic orientation that is set by the underlying electronic structure of the crystal.

\begin{figure}
    \centering
    \scalebox{1.0}{\includegraphics{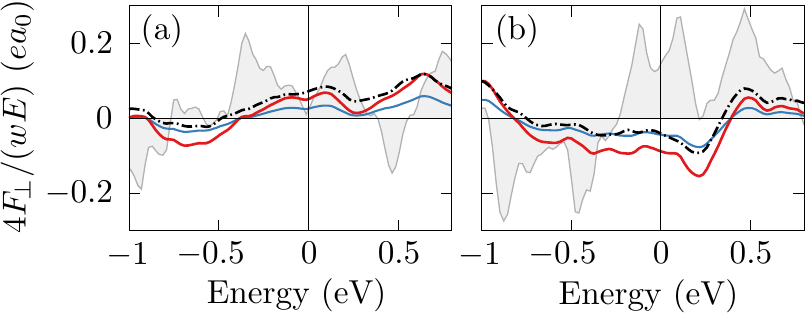}}
    \caption{Deformations of the skyrmion profile activate additional contributions to the perpendicular force $F_\perp$ in (a)~Ir/Co/Pt and (b)~Au/Co/Pt. The shaded area denotes the force for the rigid symmetric texture, whereas solid blue ($\delta c=0.1 w$) and solid red ($\delta c=0.2w$) lines represent the corresponding changes owing to deformations of the spin profile via the modulated core size $c + \delta c \sin^2\phi$. In the case of $\delta c=0.2w$, the dashed black line indicates the new force component due to the lowest-order field-like torque.}
    \label{fig:deformations}
\end{figure}

To elucidate more clearly the overall relevance of the proposed mechanism based on anisotropic field-like torques, we consider the effect of deformations of the skyrmion texture on the generalized forces. For this purpose, we assume a deformed spin structure by replacing the constant core size $c$ in Eq.~\eqref{eq:profile} with $c+\delta c \sin^2 \phi$, where $\delta c$ encodes the degree of deviation from the symmetric profile, which is recovered for $\delta c=0$. While any finite deformation of this type indeed activates perpendicular forces mediated by the lowest-order effective field $\vn H_\ut{eff}^\ut{FL} \propto \hat{\vn e}_y$, rather strong modifications of the spin structure are necessary to change $F_\perp$ considerably as shown in Fig.~\ref{fig:deformations}. However, the additional lowest-order field-like torque hardly alters the dynamics of topological spin textures if only small to medium deformations are present, in the case of which $F_\perp$ originates primarily from anisotropic field-like SOTs. At the same time, we note that the force $F_\parallel$ due to the antidamping torques remains essentially unchanged for the considered type of deformations of the skyrmion structure.

\subsection{Skyrmion Hall effect}
Using our microscopic insights into the generalized forces and the fundamental effective SOT fields, we address now the question how the higher-order torques affect the dynamical properties of N\'eel-type skyrmions. Owing to the interplay of gyromagnetic coupling, dissipation, and generalized forces in the Thiele equation~\eqref{eq:thiele}, topological spin textures usually move under an angle with respect to the applied electric field. As the magnetic analogue of the Magnus effect, this so-called skyrmion Hall effect stems from the complex spin topology in real space, and it is quantified by the skyrmion Hall angle $\theta_\ut{sk}= \arctan v_\perp / v_\parallel$. Since the dissipation tensor, Eq.~\eqref{eq:dissipation}, amounts to the unit matrix times a constant $D=\mathcal D_{xx}=\mathcal D_{yy}$ for the considered symmetric magnetic texture, we arrive at
\begin{equation}
    \frac{v_\perp}{v_\parallel} = \frac{Q F_\parallel +  D F_\perp}{ D F_\parallel - Q F_\perp} = \frac{Q}{D} + \frac{(D+Q^2/D)F_\perp}{DF_\parallel-QF_\perp} \, .
\end{equation}
Therefore, while $\theta_\ut{sk}$ is determined solely by $Q/D$ in the absence of any transverse torque-related force, in fact, the second equality emphasizes that the skyrmion Hall angle acquires a correction due to the force $F_\perp$ that can be activated either by the presence of higher-order field-like SOTs (see Fig.~\ref{fig:forces}), or by deformations of the spin structure.

\begin{figure}
    \centering
    \scalebox{0.18}{\includegraphics{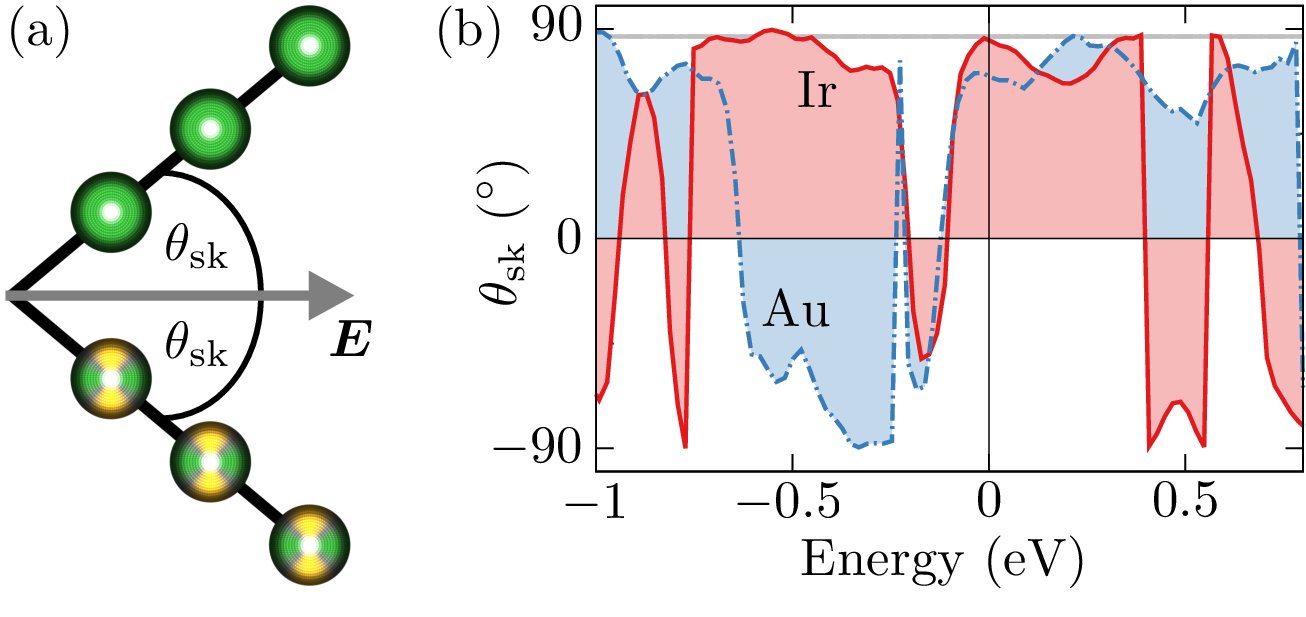}}
    \caption{(a)~Due to the skyrmion Hall effect, skyrmions and antiskyrmions move under the angle $\theta_\ut{sk}$ with respect to the applied electric field $\vn E$. The color scale for the spin textures follows the one of Fig.~\ref{fig:forces}. (b)~Skyrmion Hall angle $\theta_\ut{sk}$ in the magnetic trilayers Ir/Co/Pt (solid red line) and Au/Co/Pt (dashed blue line) as the Fermi energy is varied. The results are obtained by considering the symmetry-allowed first-principles shape of the spin-orbit torques. For comparison, the thin gray line indicates the value of $\theta_\ut{sk}$ in the absence of anisotropic field-like torques. The chosen Gilbert damping parameter of $\alpha_\ut{G}=0.05$ is assumed to be independent of the energy.}
    \label{fig:dynamics}
\end{figure}

Figure~\ref{fig:dynamics} presents the computed energy dependence of the skyrmion Hall angle $\theta_\ut{sk}$ in the considered magnetic trilayers Ir/Co/Pt and Au/Co/Pt, for which we assume a constant Gilbert parameter of $\alpha_\ut{G}=0.05$ as the Ir/Co and Au/Co interfaces exhibit similar damping characteristics~\cite{Azzawi2016,Kim2016}. Our calculations show that while $\theta_\ut{sk}$ is generally finite, which corresponds to a deflection of the symmetric skyrmion from the current line, proper engineering of the electronic structure facilitates in principle a straight motion of the spin structure without suffering from any skyrmion Hall effect. This is possible as the transverse force component that is triggered by higher-order field-like SOTs adds crucial flexibility for balancing the different terms in the Thiele equation of motion. For example, doping Au with Pt or substituting a small portion of Ir with Os can thus drastically change the skyrmion Hall angle, see Fig.~\ref{fig:dynamics}(b), as compared to the case without anisotropic torques, where $\theta_\ut{sk}$ relates to $Q/D$.

Finally, we remark that our conclusions apply analogously to antiskyrmions~\cite{Hoffmann2017,Nayak2017}, which can be obtained by reflecting the magnetization profile~\eqref{eq:profile} at a plane perpendicular to the film. Specifically, the symmetric N\'eel texture shown in Fig.~\ref{fig:torques}(b) transforms into the spin structure of an antiskyrmion if the corresponding magnetization field is mirrored at the $y=0$ plane, see Fig.~\ref{fig:torques}(c). Thus, in the rigid body approximation, if the N\'eel skyrmion moves under the angle $\theta_\ut{sk}$, the antiskyrmion exhibits the very same Hall angle but with respect to an accordingly mirrored axis as illustrated in Fig.~\ref{fig:dynamics}(a), although the dynamics of antiskyrmions can generally be more complex than for skyrmions~\cite{Ritzmann2018}.

\section{Conclusions}\label{sec:conclusions}

Using material-specific electronic-structure calculations in combination with symmetry considerations for the effective magnetic fields, we demonstrated that the anisotropy of spin-orbit torques (SOTs) is vital in understanding and predicting the dynamical properties of topological spin textures. In particular, we uncovered that higher orders of the field-like SOTs manifest in qualitatively new forces that act transverse to an applied electric field. These anisotropic field-like torques couple to the motion of radially symmetric N\'eel-type skyrmions and antiskyrmions, and imprint on their dynamics, granting thereby additional flexibility in tuning the skyrmion Hall angle in real multilayer materials. Considering as specific examples the film systems Ir/Co/Pt and Au/Co/Pt, we showed that substitutional doping of the overlayers with other heavy metals provides indeed a suitable means for engineering the dynamics of magnetic textures. 

Our findings open up an alternative path for controlling the motion of skyrmions by proper materials design of anisotropic SOTs. The predicted phenomenon competes with the conventional mechanism where the isotropic field $\vn H_\ut{eff}^\ut{FL}\propto \hat{\vn e}_y$ can effectively couple to the spin structure due to deformations of the texture. Developing a coherent picture of the interplay of these two competing effects beyond the limited treatment of the Thiele equation, for example, by atomistic spin dynamics simulations, presents an exciting research direction. However, this necessitates the extension of such simulations to incorporate the relevant effective SOT fields that are of higher order in the local magnetization. Overall, our microscopic insights into the dynamical properties of skyrmions and antiskyrmions call for reviewing the common perception that the skyrmion Hall effect is independent of the field-like SOT. Specifically, this phenomenon could acquire large corrections in systems with reduced symmetry owing to the proposed mechanism that roots in the prominent anisotropy of field-like torques.

\begin{acknowledgments}
We thank Markus~Hoffmann for insightful discussions, and we acknowledge support by the Deutsche Forschungsgemeinschaft (DFG) through Priority Programm SPP 2137 and the Collaborative Research Center SFB/TRR 173. We also gratefully acknowledge the J\"ulich Supercomputing Centre and RWTH Aachen University for providing computational resources under project jiff40.
\end{acknowledgments}

\bibliographystyle{apsrev4-2}
\bibliography{thebib}

\end{document}